\documentclass[prl,superscriptaddress,showpacs,twocolumn]{revtex4}
\usepackage{graphicx}
\usepackage{times}
\usepackage{color}
\usepackage[pdftex,colorlinks,bookmarks=false,citecolor=blue,linkcolor=red,urlcolor=blue]{hyperref}
\def\bea{\begin{eqnarray}}
\def\eea{\end{eqnarray}}
\def\be{\begin{equation}}
\def\ee{\end{equation}}
\def\kag{kagome }
\def\S{\mbox{\bf S}}

\begin{document}
\author {Andreas~M.~L\"auchli}
\affiliation{
Max Planck Institut f\"ur Physik komplexer Systeme, N\"othnitzerstr.~38, D-01187 Dresden, Germany.
}
\author{Claire Lhuillier}
\affiliation{Laboratoire de Physique Th\'eorique de la Mati\`ere Condens\'ee, UMR 7600 CNRS,
Universit\'e Pierre-et-Marie-Curie, Paris 6, 75252 Paris cedex 05, France.
}
\date{\today}
\title{Dynamical Correlations of the Kagome $S=1/2$ Heisenberg Quantum Antiferromagnet}
\begin{abstract}
We determine dynamical response functions of the $S=1/2$ Heisenberg quantum antiferromagnet on the \kag lattice
based on large-scale exact diagonalizations combined with a continued fraction technique. The dynamical spin 
structure factor has important spectral weight predominantly along the boundary of the extended Brillouin zone 
and energy scans reveal broad response extending over a range of $2 \sim 3J $ concomitant with pronounced
intensity at lowest available energies. Dispersive features are largely absent. 
Dynamical singlet correlations -- which are relevant for inelastic light probes -- reveal a similar broad response, with
a high intensity at low frequencies $\omega/J \lesssim 0.2J$. These low energy singlet excitations do however not seem 
to favor a specific valence bond crystal, but instead spread over many symmetry allowed eigenstates.
\end{abstract}
\maketitle
\paragraph{Introduction}
The $S=1/2$ Heisenberg quantum antiferromagnet (AFM) on the \kag lattice is a key model
for our theoretical understanding of highly frustrated quantum magnets in two spatial dimensions.
In contrast to other model systems, such as the checkerboard magnet, it continues to hide
the true nature of its ground state despite a longstanding effort. On the experimental side many \kag like 
materials were discovered and characterized over the years, 
and some of them seem to get close to the goal of an experimental realization of a perfect 
Heisenberg $S=1/2$ \kag system. Recently the synthesis of powder samples of the Herbertsmithite 
ZnCu$_3$(OH)$_6$Cl$_2$~\cite{HSmithite06} and the subsequent experimental investigations sparked 
a new wave of theoretical interest in this long standing problem~\cite{PhysicsToday}.

So far most of the theoretical studies focused on ground state properties, discussing various
possible phases, such as valence bond solids \cite{VBSKagome}, gapped spin liquids of different 
kinds~\cite{Sachdev92_Wang06,Mila98,DMRGJiang},
and also stable critical phases~\cite{Hastings,ASL}.
Much less attention however has been paid to the precise nature and form of the low-energy excitations
visible in frequency resolved probes such as inelastic neutron scattering or light scattering techniques.
In the present Letter we fill this void and present a detailed numerical study of the
dynamical response of $S=1/2$ \kag systems in the spin triplet and singlet channels
and formulate predictions to be tested in inelastic scattering experiments.

We study the $S=1/2$ Heisenberg quantum antiferromagnet on
the \kag lattice, governed by the Hamiltonian:
\be
H=J \sum_{\langle i,j \rangle} \S_i \cdot \S_j 
\ee
where $J>0$ is the antiferromagnetic nearest neighbor exchange coupling.
Our results are based on large-scale exact diagonalizations of up to $N=36$ spins, supplemented
by the continued fraction method~\cite{Gagliano86} for dynamical correlations functions. For the
dynamical quantities we performed 500 (3000) iterations in Hilbert spaces of dimensions $3.8 \times 10^8$
($4.5\times 10^9$) for the spin and singlet dynamics respectively.

A lot of our present understanding of $S=1/2$ systems is based on a series of exact diagonalization
studies \cite{ED_Kagome},
which convincingly showed the absence of
magnetic order, and revealed a puzzlingly high density of low-energy singlet and triplet excitations, 
a so far unique phenomenon. In the following we explore how the huge density of low-lying excitations
affects experimentally relevant response functions.

\paragraph{Static spin response}

The static spin structure factor is given by the following expression:
\bea
S^z(\mathbf{Q})&\equiv&\frac{1}{\sqrt{N}} \sum_{j}  \mbox{e}^{- i \mathbf{Q}\cdot\mathbf{r}_j}\ \S^z_j\ , \nonumber \\
\mathcal{S}(\mathbf{Q})&=&\langle S^z(\mathbf{-Q}) S^z(\mathbf{Q}) \rangle\ ,
\eea
where the wave vector $\mathbf{Q}$ is not restricted to lie in the first Brillouin zone (BZ).

\begin{figure*}
\centerline{\includegraphics[width=0.95\linewidth]{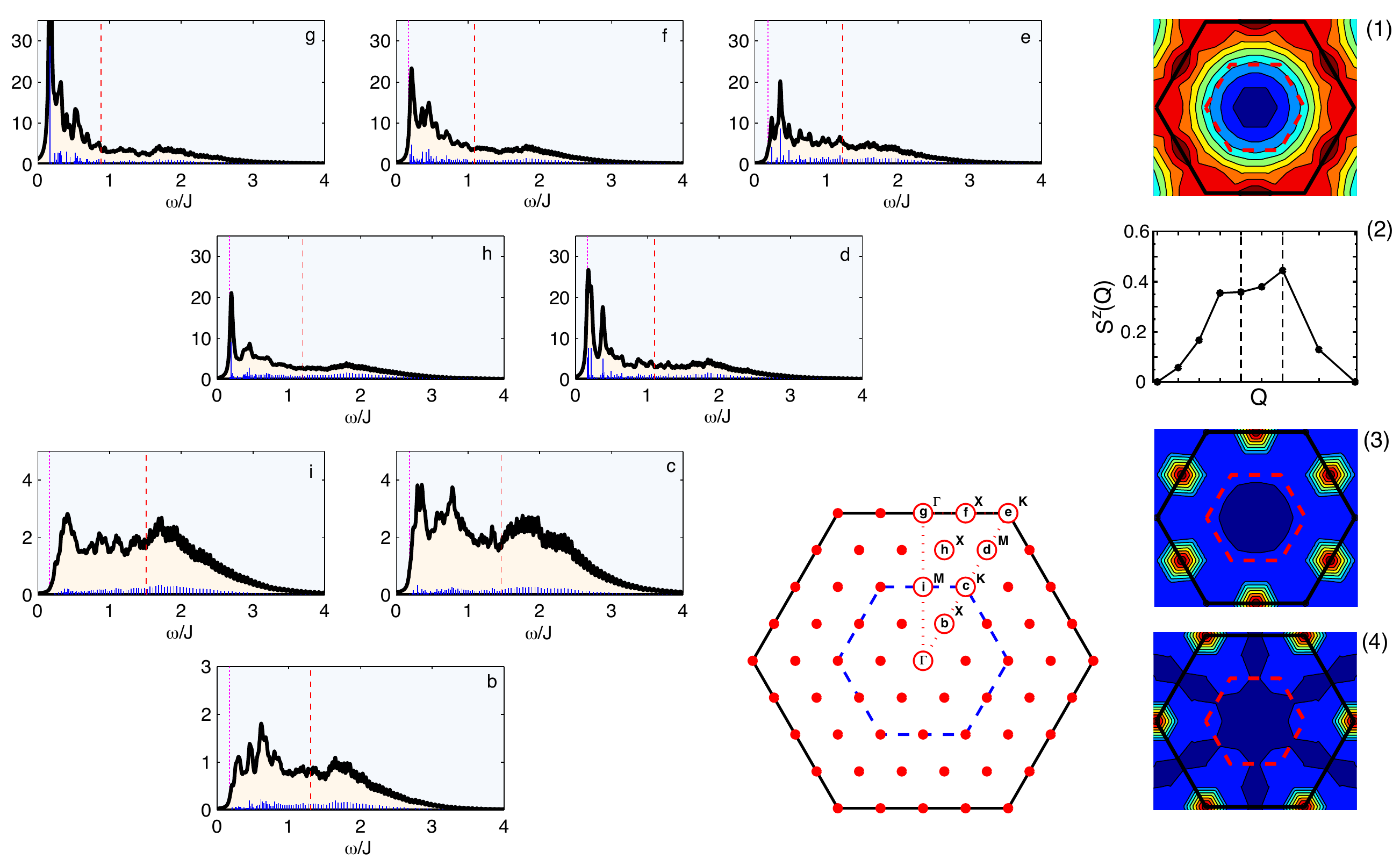}}
\caption{
(Color) Dynamical spin structure factor of the $N$=36 sample.
The eight panels display frequency scans $\mathcal{S}(\mathbf{Q},\omega)$ ($\eta=0.02J$)
at labeled wavevectors $\mathbf{Q}$ in the extended Brillouin zone shown in the
lower right center. Note that the intensity scales differ among the different panels.
The $\Gamma$ point has no weight and is not shown. The blue vertical lines show
the pole location and intensity of the continued fraction.
The vertical dotted magenta line denotes the finite size spin gap in the corresponding momentum
sector. The dashed red line marks the position of the first frequency moment
$\bar{\omega}=\int \mathrm{d}\omega\ \omega\ \mathcal{S}(\mathbf{Q},\omega)
/ \mathcal{S}(\mathbf{Q})$. In the rightmost column the static spin structure factor of the pure Heisenberg model on the kagome lattice 
is shown, as an intensity plot (1) and along the path $\Gamma-(e)-(g)-\Gamma$ (2). The static structure factor for the $q=0$ (3) and $\sqrt 3\times \sqrt 3$  (4) N\'eel order states induced by  appropriate second neighbor couplings are also displayed.
\label{fig:dynstatSq}}
\end{figure*}

The numerically determined static spin structure factor for the $36$ sites sample is shown in Fig.~\ref{fig:dynstatSq},
as an intensity plot covering multiple BZs [right hand side, plot (1)], and along a path in the extended BZ [plot (2)].
The response is strong and broad along the zone boundary of the extended BZ, revealing the short ranged nature of the
antiferromagnetic correlations~\cite{ED_Kagome}. In the $N=36$ groundstate there are additional small peaks at
the point $(g)$, a feature which is also reported in a recent DMRG study~\cite{DMRGJiang}. Comparison with a $N=24$ sample
(not shown) confirms that the broad response along the boundary of the exended BZ is a generic, size-independent feature,
characterizing a state with spin correlations which are decaying rapidly beyond the nearest neighbor sites.
To contrast this result with a magnetically ordered state we show the static response in the $\mathbf{q}=0$ [Fig.~\ref{fig:dynstatSq}(3)]
and $\sqrt{3}\times\sqrt{3}$  [Fig.~\ref{fig:dynstatSq}(4)] states, which have been obtained using a strong $J_2$ coupling with the
appropriate sign.

{\em Dynamical spin structure factor} The energy and momentum dependence of the dynamical structure factor:
\be
\mathcal{S}(\mathbf{Q},\omega)=-\frac{1}{\pi} \mathrm{Im}
 \langle S^z(\mathbf{-Q})
\frac{1}{\omega-(H-E_\mathrm{GS}) +i\eta} S^z(\mathbf{Q}) \rangle\ ,
\ee
is directly relevant for inelastic neutron scattering (INS) experiments and therefore a quantity of central interest. In magnetically ordered systems
we expect to see dispersive, long-lived spin waves~\cite{ChristensenPNAS}, while one-dimensional systems in appropriate regimes
can display spinon continua with a rich structure~\cite{LakeNatMat}.

Our numerical results for the $N=36$ \kag lattice are presented in the left part of Fig.~\ref{fig:dynstatSq}. The shaded panels display an energy
scan at the wave vector indicated by the panel position and its label, referring to specific points in the extended BZ. 
Each panel displays the broadened ($\eta{=}0.02J$) spectral function (black line), the locations and weights of the poles of the continued fraction expansion (blue vertical lines), the finite size spin gap in the corresponding momentum sector (dotted vertical line), and the first frequency moment
$\bar{\omega}(\mathbf{Q})=\int \mathrm{d}\omega\ \omega\ \mathcal{S}(\mathbf{Q},\omega) / \mathcal{S}(\mathbf{Q})$ (dashed vertical line).

Consistent with the static structure factor [by virtue of  the sum rule $\mathcal{S}(\mathbf{Q})=\int \mathrm{d}\omega\ \mathcal{S}(\mathbf{Q},\omega)$], the dynamical spin response function concentrates essentially along the boundary of the extended BZ. 
The main feature of this system is the stretching of the magnetic response in each ${\bf Q}$-sector on a very large number of excited states spanning a large bandwidth of $2 \sim 3J$,  starting immediately above the (finite-size) gap. Furthermore there seems to be a pronounced enhancement of the intensity at small $\omega$.
The different spectral functions look rather similar, suggesting an approximate factorization $\mathcal{S}(\mathbf{Q},\omega) \sim \mathcal{S}(\mathbf{Q}) \times f(\omega)$, at least at intermediate and high $\omega$. The overall picture is definitely quite different from the spectrum of a N\'eel ordered system on the same system size, where an overwhelming part of the spectral weight is carried by very few poles in each ${\bf Q}$-sector associated to the Bragg peak and the one-magnon modes respectively~\cite{LuscherHBAFM}. Still at some wave
vectors the lowest pole carries significant weight [especially at $(g)$]. The origin of this feature remains to be elucidated but could potentially come from an algebraic divergence in one scenario~\cite{ASL} or from triplon excitations on top of (remnants of) a valence bond crystal in a different scenario~\cite{VBSKagomeExcitations}.

\begin{figure}
\centerline{\includegraphics[width=0.9\linewidth]{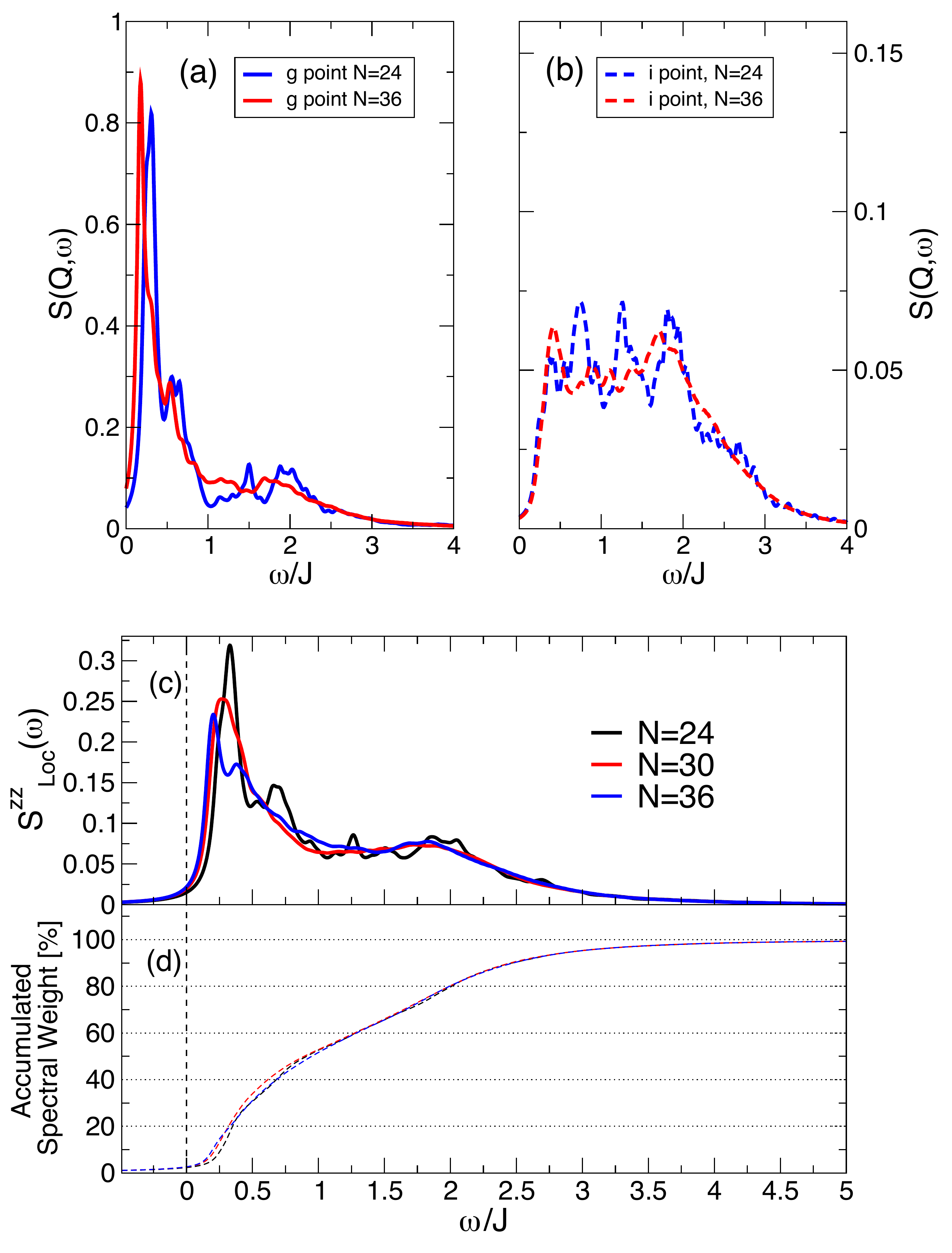}}
\caption{
(Color online) {\em Upper panel:} Finite size behavior of the spectral functions at two
different points in the Brillouin zone: $g$ (a) and $i$ (b) in the convention of
Fig.~\protect{\ref{fig:dynstatSq}}.
{\em Lower panel:}
(c) Local spin autocorrelation function
$\mathcal{S}_\mathrm{loc}(\omega)\propto \int d\mathbf{Q}\ \mathcal{S}(\mathbf{Q},\omega)$
for $N= $ 24, 30 and 36 sites.
(d) The cumulative spectral weight as a function of $\omega/J$.
All spectral functions have been broadened using $\eta=0.05J$.
\label{fig:localSofq}}
\end{figure}
In order to address finite-size effects we present two spectral functions at the wave vectors $(g)$ and $(i)$ for $N=24$ and $36$ spins in Fig.~\ref{fig:localSofq}(a) and (b). The characteristic width in energy as well as the prominent response at low $\omega$ for wave vector $(g)$ are clearly stable with respect to finite size effects.
In panels (c) the local dynamical spin correlation function $\mathcal{S}_\mathrm{loc}(\omega)\propto \int d\mathbf{Q}\ \mathcal{S}(\mathbf{Q},\omega)$  is shown for
different system sizes, again highlighting the stability of the overall shape of the spectral function. Finally panel (d) presents the cumulative spectral weight as a function of $\omega/J$, revealing e.g. that $\omega \sim 0.95 J$ is the median frequency for the total spin response.

\paragraph{Effect of impurities} We have studied the influence of a low concentration of nonmagnetic impurities on the spin dynamics
by depleting a $N=27$ sample by one site. The averaged dynamical spin response closely resembles Fig~\ref{fig:localSofq}(c)
with an additional resonance-like feature at $\omega\sim J$ due to the strong singlet forming on the two bonds
next to the vacant site~\cite{DommangeOneHalfImpurity}. Due to its local nature, this feature is expected to be generically present in 
$\mathcal{S}(\mathbf{Q},\omega)$ as well.

\begin{figure}
\includegraphics[width=\linewidth]{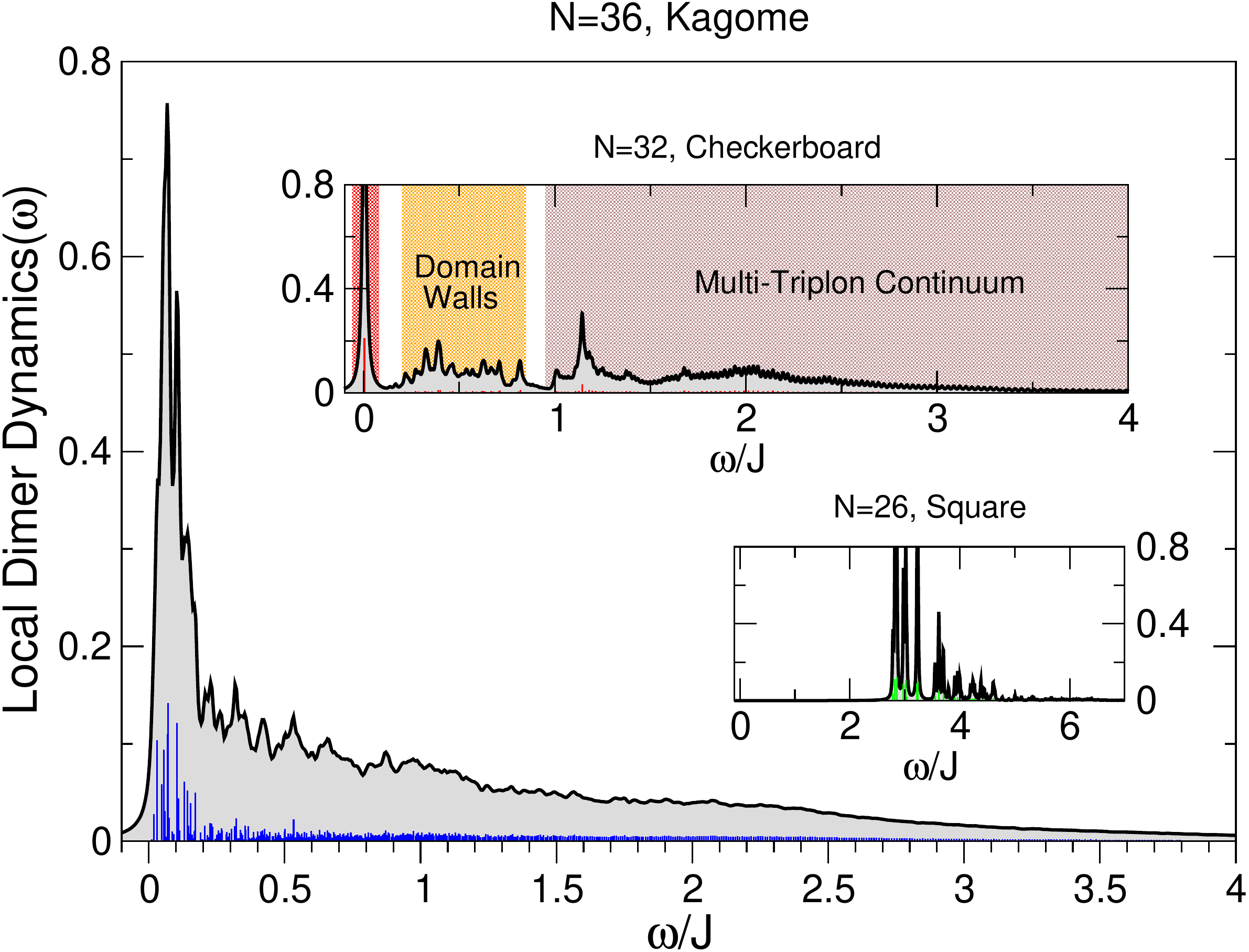}
\caption{
(Color) Dynamical singlet fluctuations [Eq.~(\protect{\ref{eqn:dimerfluct}})] for three different systems.
{\em Main plot:} \kag lattice. {\em Upper inset:} Checkerboard lattice with a plaquette-like valence bond crystal ground state.
{\em Lower inset:} Unfrustrated square lattice exhibiting N\'eel order.
The plotted quantity represents well the qualitative features of the Raman response  of the three
systems.
\label{fig:dimerfluctuations}}
\end{figure}
\paragraph{Singlet fluctuations}
In order to assess the importance of the abundant number of low energy singlet excitations
for optical probes and to investigate the tendency towards valence bond crystal ordering,
we study the local dynamical fluctuations of a nearest neighbor dimer operator:
\bea
D_{i,j}&=&\S_i\cdot\S_j - \langle \S_i\cdot\S_j \rangle \nonumber \\
\label{eqn:dimerfluct}
\mathcal{D}_{i,j}(\omega)&=&-\frac{1}{\pi} \mathrm{Im}\langle D_{i,j}  \frac{1}{\omega-(H-E_\mathrm{GS})+i\eta} D_{i,j} \rangle
\eea
The interest in this quantity is twofold. First Eq.~(\ref{eqn:dimerfluct}) is closely related to the Raman or RIXS reponse of a spin 
system
and thus reveals the qualitative features of the inelastic light scattering response.
And second, we expect a spontaneous translational symmetry breaking due to dimerization to show up as an important $\omega \to 0$ contribution.

The fluctuation spectrum for the $N=36$ \kag system is shown in Fig.~\ref{fig:dimerfluctuations}, where a broad response
from the lowest singlet up to energies $\sim 4 J$ is seen together with a strong increase of the response towards the lowest energies. 
The \kag result can be compared to the response of the Heisenberg model on a checkerboard lattice (upper inset in Fig.~\ref{fig:dimerfluctuations}) 
and the square lattice (lower inset in Fig.~\ref{fig:dimerfluctuations}), where in both cases the physical origin of the response is essentially understood.
On the N\'eel ordered square lattice the dimer-dimer response is assigned to the two-magnon continuum with a maximum strength around $3J$. This is obviously very similar to the Raman response of the square lattice Heisenberg AFM \cite{Sandvik98}.  On the checkerboard lattice with its plaquette valence bond crystal ground state~\cite{FouetPyro} the dimer-dimer response function shows different frequency domains: a single low-lying peak (shaded in red) originating from the valence-bond symmetry breaking partner of the ground state, followed by two domains at non zero-frequency, first a range of singlet excited levels (shaded orange) which have been convincingly explained as valence bond crystal domain-wall excitations~\cite{BergPyro,FouetPyro} and a second range $\omega \gtrsim J$ corresponding to a multi-triplon continuum~\cite{FouetPyro} (shaded in brown).
It is clear from this comparison that the \kag lattice does not show typical valence bond crystal characteristics exemplified by the checkerboard magnet. Still 
the response on several levels up to $\omega\lesssim 0.2J$ seems to be particular strong. However this response spreads on many low lying levels in any symmetry sector which can be excited by the dimer-dimer operator. We do not see any clear precursor of a specific spatial symmetry breaking pattern when comparing the excited levels to different valence bond crystal symmetry predictions summarized in Ref.~\cite{MisguichVBCSymmetry}. Possible reasons are: i) the absence of VBC order (see
Ref.~\cite{DMRGJiang} for a similar conclusion) or
ii) a very weak ordering with a very large unit cell which has not still emerged from competing orders.

\paragraph{Comparison to theoretical proposals}

Both the spin and the dimer dynamical fluctuations of a $S=1/2$ \kag system are intrinsically broad, and are not easy to reconcile with the
excitation spectrum scenarios for the various proposed ground states. For example the dynamical spin correlation functions lack the characteristic
coherent triplon excitations, which are expected on top of a valence bond crystal~\cite{VBSKagomeExcitations}. On the other hand the spectrum is much too dense to be explained
by a spinon continuum with a energy scale of order $J$~\cite{Sachdev92_Wang06}. A critical spin liquid would possibly have similar low-$\omega$ enhanced
response~\cite{ASL} as we report, but the current theoretical predictions would still predict dispersive structures with a bandwidth significantly larger than our numerical
bandwidth.
Interestingly our fully quantum mechanical results show some similarities with fluctuation spectra of classical highly frustrated systems~\cite{ClassicalHFMDynamics}, 
where the macroscopic ground state degeneracy plays an important role. It is thus possible that the intermediate and high energy response of the $T=0$ 
\kag $S=1/2$ quantum magnet resembles that of a classical cooperative paramagnet at finite temperature, while the precise low-energy response will ultimately be dictated by the yet unknown true nature of the ground state.

\paragraph{Comparison to experiments}

Our results provide definite predictions for the spectral response of a perfect $S=1/2$ \kag system that can be checked in experiments.
First of all a pure Heisenberg model on the kagome lattice has a static spin structure factor with a negligible weight inside the first BZ and most of the weight is
concentrated on the boundary of the extended BZ. While neutron diffraction data on many $S>1/2$ systems is in good agreement with this expectation,
the same data for the $S=1/2$ Herbertsmithite~\cite{Helton07,LeeNatMat} is difficult to reconcile with a simple Heisenberg model and possibly 
points to some remnant magnetism that may be due to impurities or Dyaloshinsky-Moryia interactions. 
Similarly for the inelastic response: while $S=5/2$ Deuterium-Jarosites are in nice qualitative agreement~\cite{FakDJaro07} with our results, 
the reported inelastic neutron signals for the Herbertsmithites~\cite{Helton07,LeeNatMat} is concentrated at energies lower than what we would expect for a pure Heisenberg model.
Therefore more
experimental and theoretical work is needed in order to understand the Herbertsmithite response.
Finally our results might also provide a first step towards an understanding of the anomalous dynamical spin fluctuations revealed by $\mu$sr on
SrCr$_8$Ga$_4$O$_{19}$~\cite{Uemura94} and related compounds.

\paragraph{Perspectives}

Based on dynamical correlations functions in the tripled and singlet channels we have shown that the \kag system is a highly
fluctuating magnet with broad response peaking at low energies. 
This fluctuating background provides also a natural explanation for the reported absence of quasiparticles in a
doped \kag system~\cite{AMLSpinCharge}.
Building on these results it will be interesting to understand the evolution of the dynamical response at finite temperature
as well as the effect of a magnetic field. Given that many experimental \kag systems have some (small) Dzyaloshinsky-Moriya 
interactions their influence on the dynamical response is also worth studying.

\acknowledgments

We thank C.L.~Henley, F.~Mila, R.~Moessner, D.~Poilblanc and K.P.~Schmidt for discussions
and comments, and M. de Vries, G. Nilsen and H.M. R\o nnow for discussions and
access to their unpublished neutron scattering data. We acknowledge support
by the Swiss National Funds. The computations have been performed on the machines of the
MPG RZ Garching and the CSCS Manno.

\end{document}